\documentclass[a4paper]{article}
\usepackage[affil-it]{authblk}

\usepackage{graphicx}

\usepackage{color}

\title{Thermoelectric effects in nanostructured quantum wires in the non-linear
temperature regime}
 
\author[1]{Anda Elena Stanciu}
\author[1,2]{G. A. Nemnes}
\author[3]{A. Manolescu}
\affil[1]{University of Bucharest, Faculty of Physics, Materials and Devices for Electronics and Optoelectronics Research Center, 077125 Magurele-Ilfov, Romania}
\affil[2]{Horia Hulubei National Institute for Physics and Nuclear Engineering, 077126 Magurele-Ilfov, Romania}
\affil[3]{School of Science and Engineering, Reykjavik University, Menntavegur 1, IS-101 Reykjavik, Iceland}

\date{}

\begin{document}
\maketitle
\begin{abstract}
The thermoelectric voltage of a quantum dot connected to leads is
calculated using the scattering R-matrix method.  Our approach takes into
account a temperature gradient between the contacts beyond the
linear regime.  We obtain sign
changes of the thermopower when varying the temperature or the chemical
potential around the resonances.  The influence of the coupling strength
of the contacts and of the thermoelectric field on the thermoelectric
voltage is discussed.\\
{\em Keywords}: thermoelectrics, R-matrix, non-linear temperature regime.
\end{abstract}

\section{INTRODUCTION}

The history of nanoscale thermoelectrics dates back to the 90's when it was shown that quantum wires and quantum wells may provide an enhancement of the figure of merit compared to the calculated value for the bulk structures \cite{1}. Since then, studies on various quantum systems have been performed, and especially on thermoelectric 
efficiency \cite{humprey,sothmann,dwyer,karlstrom}.
Small diameter nanowires \cite{heremans,nemnes1} as well as
quantum dot systems \cite{svensson} investigated from both theoretical and experimental point of view provide special thermoelectric characteristics as they present sharp features \cite{mahan} in the transmission functions.
The nonlinear temperature regime offers new ways to obtain an enhanced thermoelectric efficiency. 
In this study we analyze thermoelectric effects which arise due to a temperature
difference between the contacts of a device model using the scattering formalism based on the R-matrix. We present the typical features
observed in resonant systems \cite{jianming}, such as the sign-change of the thermoelectric voltage by
changing the chemical potential around and in-between the resonances \cite{Beenakker,KT}, but also by
varying the temperature of the contacts. 
The influence of the coupling between the quantum system
and the contacts is analyzed. 
Finally, the non-linear effects introduced by the thermoelectric field between the contacts are 
incorporated and the possible enhancement of the thermoelectric voltage is discussed.

\section{Model and method}

\begin{figure}[h]
\centering
\includegraphics[width=6.0cm]{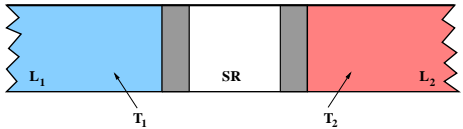}
\caption{Schematic representation of the two-dimensional system composed by the two leads and the scattering region SR. The gray areas are potential barriers which describe the couplings between SR and the 
two leads L$_{1,2}$ which have different temperatures, $T_2>T_1$.}
\label{Figure 1}
\end{figure}

The two-terminal model device is a double barrier system as illustrated in Fig. \ref{Figure 1}.
The leads are invariant along the transport direction and the perpendicular confinement is
parabolic, with a level spacing $h\omega_{0} = 100$ meV.
The contacts have distinct temperatures $T_1$ and $T_2$.  

In the framework of multi-channel coherent transport \cite{nemnes1,nemnes2,nemnes3}, the transmission function is calculated as  
\begin{equation}
{\mathcal T_{ss'}}(E)=\sum_{i, i'} \Theta \left(E - E^{\nu}_{\perp} \right)\Theta \left(E - E^{\nu'}_{\perp} \right) {\left| \ \tilde{S}_{\nu\nu'}\left(E\right) \right|}^{2} \; , 
\label{Tss'}
\end{equation}
where $\nu=(s,i)$ denotes the channel index $i$ in lead $s\ (s=1,2)$   
and $\tilde{S} = K^{\frac{1}{2}} S K^{-\frac{1}{2}}$. 
The diagonal matrix $K$ is formed by longitudinal wavevectors,
$k_\nu=\sqrt{2m(E-E^{\nu}_{\perp})}/\hbar$, where $E^{\nu}_{\perp}$
is the transversal energy for channel $\nu$ and $E$ is the total energy.
The scattering S-matrix is found in the R-matrix method as
$S =-(1+\frac{i}{m^{*}}RK)/(1-\frac{i}{m^{*}}RK)$.
By using the Heaviside functions $\Theta$ in Eq.\ (\ref{Tss'}) only the open channels are included.
The current is calculated as:
\begin{equation}
I \equiv I_{ss'} = \frac{2e}{h} \int dE \; {\mathcal T_{ss'}}(E) \; \left[ f_{FD}\left(E; \mu_1, T_1\right)-f_{FD}\left(E; \mu_2, T_2\right) \right]  \; ,
\end{equation}
where $f_{FD}(E;\mu,T)$ is the Fermi-Dirac distribution function.
The open circuit thermoelectric voltage $U_{th}$ produced by the temperature difference 
$\Delta T = T_2 - T_1$ is found
by imposing a vanishing current condition, $I=0$.
The value $U_{th}$ is determined iteratively, by changing step-wise the relative positions of the chemical potentials, which in turn modify the 
internal electric field between the contacts and the scattering potential as well.     

\vspace*{-0.5cm}
\section{Results}

We use rectangular barriers of height $V_b = 200$ meV and width $w=4.3$ nm.
The distance between the contacts, including the barriers, is $L = 46$ nm. 
The effective mass of the electron is as for InAs, $m^* = 0.023 m_e$. 
We consider $T_1=0$, but $T_2\neq 0$.

\begin{figure}[h]
\centering
\includegraphics[width=8cm]{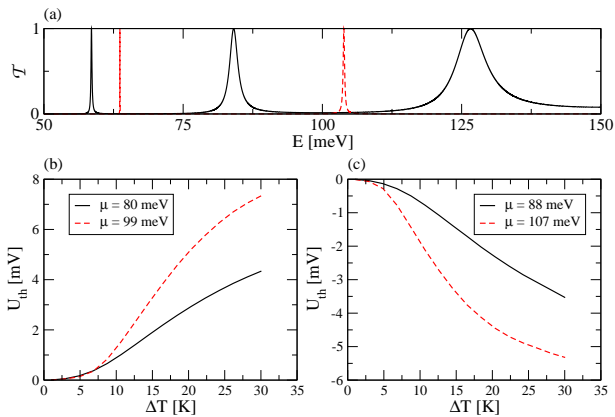}
\caption{(a) The transmission functions for systems with barrier widths $w=4.3$ nm (black/solid) and $2w$ (red/dashed). Thermoelectric voltage vs. temperature difference, with the chemical potential at zero bias $\mu$ chosen below (b) and above (c) the energy of the second resonance peak.}
\label{trans_U_T} 
\end{figure}

We first examine the influence of the coupling of the resonant system with the contacts by adjusting the widths of the  barriers. The transmission functions are depicted in Fig.\ \ref{trans_U_T}(a),
for the system with the specified barrier width $w$, and, for comparison, for $2w$.  
By doubling the barrier width the coupling to the leads weakens and the energy levels
are shifted at higher energies while the transmission peaks become narrower.
For each system we consider two values for the chemical potential, one 
placed below and one above the second resonant level $E_2^{res}$. 
The thermoelectric voltage as a function
of temperature difference is presented in Figs. \ref{trans_U_T}(b) and (c).
In order to compare the effects produced by the resonance broadening rather than energy shift, 
the initial chemical potentials of the leads, i. e. at zero bias, denoted by $\mu$, are chosen symmetrically, 
on both sides of the transmission maximum of the second resonance peak. 
The thermoelectric voltage is calculated as $U_{th}=(\mu_1-\mu_2)/|e|$, $\mu_{1,2}$ being the 
chemical potentials of the leads, where $\mu_2=\mu$ is fixed and $\mu_1$ is adjusted until the 
current generated by the temperature gradient vanishes.

Two main features are observed. First, the sign of $U_{th}$
changes as the chemical potential is situated above or under the resonant level. This indicates
that the thermoelectric current generated at zero bias has changed sign.
Second, the weaker coupling, leading to narrower transmission peaks, corresponds to
an enhanced $|U_{th}|$.  
The behavior can be explained as follows: in the limit of narrow peaks, the condition $I=0$
is fulfilled as the position of the chemical potential in the cold contact, $\mu_1$, reaches 
the resonant level, $E_2^{res}$, and the populations in the two contacts become equal; 
if the resonance is broader the vanishing current condition is reached sooner 
as $\mu_1$ approaches $E_2^{res}$, leading to smaller $|U_{th}|$.

\begin{figure}[h]
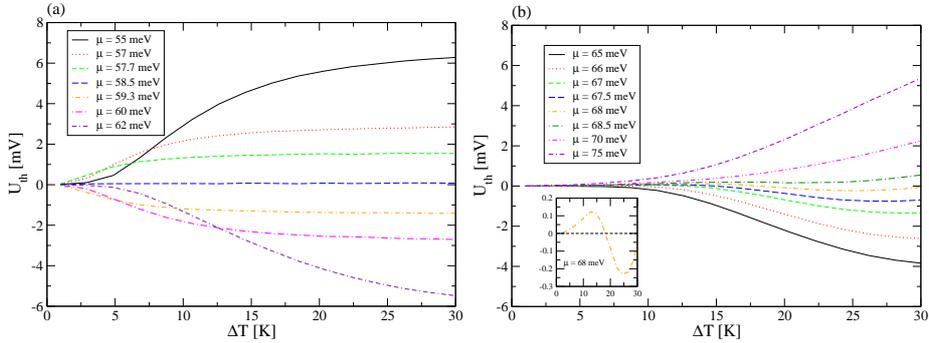

\includegraphics[width=6.0cm]{figure3a} 
\includegraphics[width=6.0cm]{figure3b}
\caption{Thermoelectric voltage vs. temperature difference: chemical potential varying around the first resonance (a) and in-between the first two resonances (b). Inset: sign change
of $U_{th}$ at a fixed $\mu$.}
\label{Figure 3}

\end{figure}

Next, for the system with barrier width of 4.3 nm the zero-bias chemical potential is varied around the first resonance and also between the first two resonances. The two resonances occur at $E=58$ and $E=84$ meV, Fig. \ref{Figure 1}(a).
The results are shown in Figs. \ref{Figure 3}(a) and (b).
In the first case the voltage curves have intersections at finite temperatures.
At temperatures below the intersections $|U_{th}|$ increases as the chemical 
potential is set closer to the resonance maximum and at higher temperatures the opposite situation
occurs. This is a consequence of the overlap of the Fermi-Dirac distribution functions with the finite
width of the resonance.  
In the second case, one also obtains a sign change for $U_{th}$ at a fixed temperature, because by changing the 
chemical potential one out of the two resonances becomes dominant. Moreover, the same effect is
also found at a fixed chemical potential and variable temperature, 
as one can see in the inset of Fig.\ \ref{Figure 3}(b).   

\begin{figure}[h]
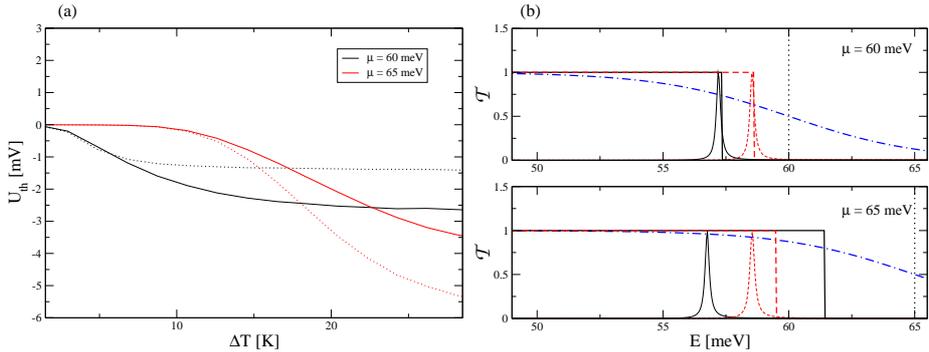

\centering
\includegraphics[width=6.cm]{figure4a}
\includegraphics[width=6.cm]{figure4b}
\caption{(a) $U_{th}$ vs. $\Delta T$ in the presence of the thermoelectric
field (solid lines) and in the absence of the thermoelectric field (dotted lines). 
(b) Transmission functions and populations in contact 1 for the case {\it with}
(black/solid) and {\it without} (red/dashed) thermoelectric fields, at $\Delta
T = 30$K. The populations in contact 2 are represented by (blue/dot-dashed) lines.
}
\label{U_T_E}
\end{figure}

The thermoelectric fields arising between the contacts 
${\mathcal E}_{th}=U_{th}/L$
constitute a source of non-linear behavior and
possible enhancement of the thermoelectric effects.  Numerical results
are shown in Fig. \ref{U_T_E}(a) to compare the
case {\it with}  and {\it without} ${\mathcal E}_{th}$ 
included in the transmission function given by Eq. \ref{Tss'}. 
The chemical potentials are selected in-between the
first two resonances for $w=4.3$ nm. 
We find that if the chemical potential is close to the resonance
the inclusion of ${\mathcal E}_{th}$ enhances $|U_{th}|$, whereas if $\mu$ is 
farther away the opposite is true.
The situation is illustrated in Fig.\ \ref{U_T_E}(b) for two values
of the chemical potential. In the presence of the
thermoelectric field the resonant levels are shifted in the same
direction as the chemical potential $\mu_1$. At $\Delta T = 30$K,
for $\mu = 60$ meV this produces an increase of $|U_{th}|$.  However,
if we set $\mu = 65$ meV, the population at the resonance is larger
and the positions of $\mu_1$ for the cases {\it with} and {\it without}
${\mathcal E}_{th}$ become reversed and $|U_{th}|$ decreases.

\vspace*{-0.5cm}
\section{Conclusions}
\vspace*{-0.2cm}
The thermoelectric effects which arise in the non-linear temperature
regime in a multi-resonance system were discussed in the framework of
coherent transport, using a scattering formalism based on the R-matrix
method.  We point out the typical features of sign changing of the
thermoelectric voltage with respect to varying the chemical potential
around a resonance and in-between two consecutive resonances.  In the
latter case, this sign change is also obtained by modifying the temperature
difference at a fixed chemical potential. This effect is shown in the recent
Ref. \cite{Sierra} in the presence of Coulomb
interaction of electrons, whereas we could obtain it without interactions.
Furthermore, we discussed the conditions for enhancing the thermoelectric
effect in the non-linear regime.\\

\vspace*{-0.3cm}
{\bf Acknowledgement}\\
Support of the EEA Financial Mechanism 2009-2014 staff mobility
grant, the hospitality of Reykjavik University, and instructive discussions with
Sigurdur Ingi Erlingsson are gratefully
acknowledged. G. A. Nemnes also acknowledges support from ANCS under
grant PN-II-ID-PCE-2011-3-0960.

\end{document}